\RequirePackage[table]{xcolor}
\documentclass{PoS}

\title{Collection and harmonization of system logs and prototypal Analytics services with the Elastic (ELK) suite at the INFN-CNAF computing centre}

\ShortTitle{Log collection and Analytics with the ELK suite at the INFN-CNAF computing centre}

\author{\speaker{T. Diotalevi}\\
        University of Bologna and INFN, Italy\\
        E-mail: \email{tommaso.diotalevi@studio.unibo.it
        }}
        
\author{A. Falabella, B.Martelli, D. Michelotto, L. Morganti\\
       INFN-CNAF, Italy}

\author{D. Bonacorsi, L. Giommi, S. Rossi Tisbeni\\
       University of Bologna and INFN, Italy}

\abstract{The distributed Grid infrastructure for High Energy Physics experiments at the Large Hadron Collider (LHC) in Geneva comprises a set of computing centres, spread all over the world, as part of the Worldwide LHC Computing Grid (WLCG). In Italy, the Tier-1 functionalities are served by the INFN-CNAF data center, which provides also computing and storage resources to more than twenty non-LHC experiments. For this reason, a high amount of logs are collected each day from various sources, which are highly heterogeneous and difficult to harmonize. In this contribution, a working implementation of a system that collects, parses and displays the log information from CNAF data sources and the investigation of a Machine Learning based predictive maintenance system, is presented.}

\FullConference{International Symposium on Grids \& Clouds 2019, ISGC2019\\
		31st March - 5th April, 2019\\
		Academia Sinica, Taipei, Taiwan}
		
\emergencystretch=\maxdimen
\begin{document}

\section{Introduction}
The Worldwide LHC Computing Grid (WLCG) is a global computing infrastructure with the mission of providing computing resources to store, distribute and analyse the data generated by the Large Hadron Collider (LHC), making the data equally available to all countries, regardless of their physical location. Computing infrastructures through which Grid services operate are hierarchically classified into \textit{Tiers} according to the kind of services they offer.\\
Since 2003, the Italian Tier-1 for the High Energy Physics experiments is hosted at the Bologna INFN-CNAF Data Center \cite{INFN:cnaf}, providing the resources, support and services needed for data storage and distribution, data processing and analysis as well as the production of simulated ('Monte Carlo') data, representing also a key computing facility for many non-LHC communities, making it one of the most important centers for distributed computing in Italy.\\
A key challenge, therefore, is the modernization of the center to be able to cope with the increasing flux of data expected in the near future e.g. with the new phase of operations of the High-Luminosity LHC \cite{LHC:hilumi}. These high-level standards of operation require a continuous work towards a full understanding of service behaviours and a constant seek for higher level of automation and optimization. Data centers worldwide are now witnessing the use of Artificial Intelligence (AI) solutions pushing them into a new phase, in which tasks traditionally managed by operators could be more efficiently managed by human-supervised machine decisions.\\
Besides, CNAF collects a large amount of logs every day from various sources, which are highly heterogeneous and difficult to harmonize: such log data is archived but almost never used, except for specific internal debugging and monitoring operations. This contribution, together with the implementation of a system that collects, parses and displays the log information from CNAF data sources, undertakes the investigation of a Machine-Learning-based predictive maintenance system, moving away from a preventive replacement of equipment, which is highly expensive and far from optimal efficiency.

\section{The Elastic Stack}
\begin{figure}
\centering
\includegraphics[width=0.8\textwidth]{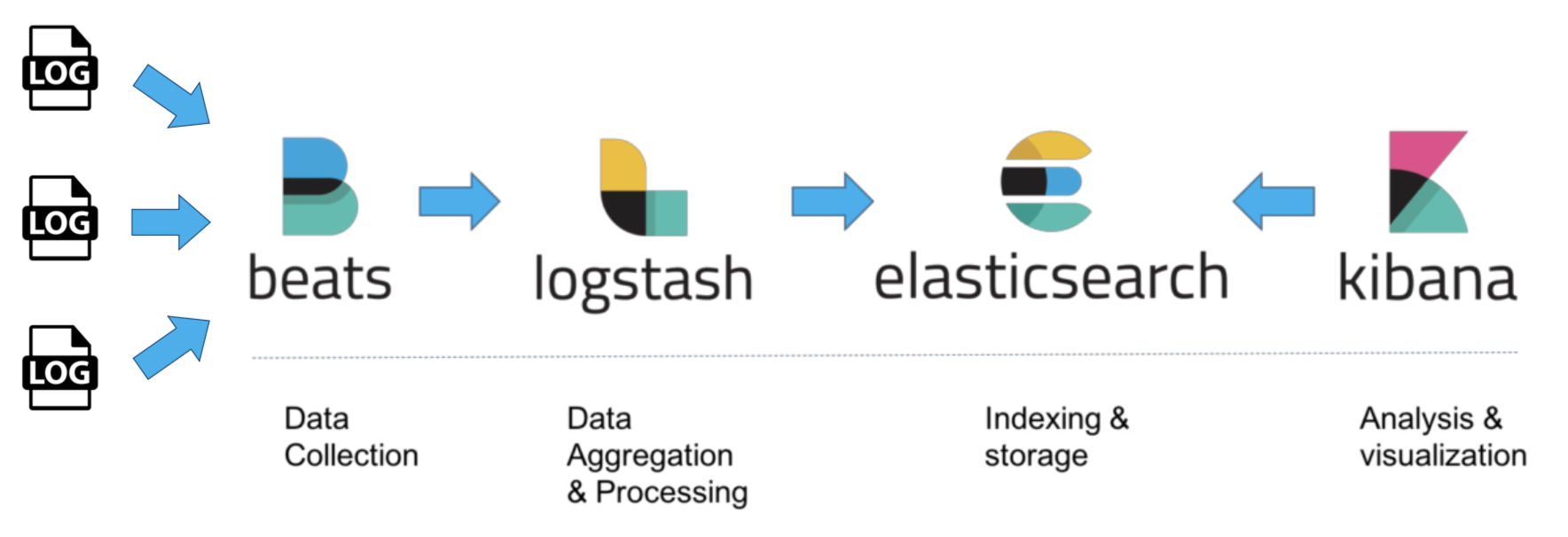}
\caption{Workflow of the Elastic Stack log ingestion and processing.}
\label{fig:elk_workflow}
\end{figure}

In order to create an indexed system with structured information from CNAF system logs, the Elastic Stack \cite{ELASTIC:elk} has been chosen. The Elastic Stack is the union of three open source project developed by the Elastic company \cite{ELASTIC:company}: Elasticsearch \cite{ELASTIC:elasticsearch}, Logstash \cite{ELASTIC:logstash} and Kibana \cite{ELASTIC:kibana}. As shown in Figure \ref{fig:elk_workflow}, logs coming from different services and different nodes are firstly collected into a single stream by a platform called Beats \cite{ELASTIC:beats}, then they are aggregated and processed by Logstash, indexed and stored via the Elasticsearch search engine and finally visualized by the Kibana user interface. In the following, a brief description of these components is given.

\subsection{Beats}
Beats is the platform that handles the shipment of data coming from different sources. Different shippers exist depending on the kind of data that needs to be moved: Filebeat for log files, Metricbeat for metrics, etc. The installation of such services occurs directly on the servers that contain data and, for this reason, is fast and lightweight. In case the information collected from Beats is unstructured, data is shipped to Logstash for further transformation and parsing.

\subsection{Logstash}
\begin{figure}
\centering
\includegraphics[width=0.7\textwidth]{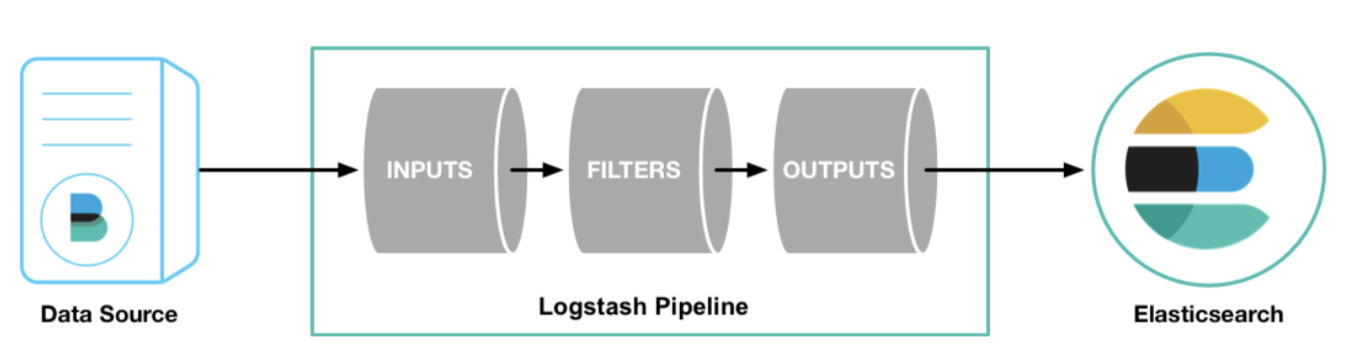}
\caption{Logstash pipeline.}
\label{fig:elk_logstash}
\end{figure}

Logstash is an open source, server-side data processing pipeline that ingests data from a multitude of sources simultaneously, transforms them, and then sends them to a central processing engine (e.g. Elasticsearch). Since data is often scattered across many systems in many formats, Logstash allows to ingest logs, metrics, web applications, etc. in a continuous, streaming fashion.\\
As data travel from source to source, as shown in Figure \ref{fig:elk_logstash}, Logstash filters parse each event, identify named fields to build structure, and transform them to converge on a common format for easier, accelerated analysis. In this regard, Logstash dynamically transforms and prepares your data regardless of its format or complexity using different filters (e.g. grok filters \cite{grok:filter} to derive structure from unstructured data, or geo filters to locate geographical coordinates from IP addresses). Data is then ready to be process by Elasticsearch, which is the embedded service provided by Elastic (this is not the only existing solution, e.g. Apache Kafka or MongoDB for logs data and InfluxDB for metrics data). 

\subsection{Elasticsearch}
Elasticsearch is the central component of the Elastic Stack with a distributed search and analytics engine capable of solving a growing number of use cases. Using a RESTful API architecture \cite{RESTful:api}, it allows to perform and combine many types of searches - structured, unstructured, geo, metric - using a Lucene querying syntax \cite{lucene:apache}. Created by Shay Banon in 2004, it was developed to provide a scalable search engine and a distributed indexing system, capable of replicating and routing data into different shards spread across the entire cluster.\\
Elasticsearch, currently, has been adopted and is running in several companies and organizations like Mozilla, GitHub, Netflix, Wikimedia, and in scientific endeavours (e.g. CERN).

\subsection{Kibana}
Kibana is an open source data visualization plugin for Elasticsearch. It provides visualization of the indexed content of an Elasticsearch cluster. Using the embedded user interface, users can create bar, line and scatter plots, or pie charts and maps based on large volumes of data. Kibana also provides several additional features, such as developer tools for advanced interactions with the Elastic Stack e.g. a console able to interact with the REST API of Elasticsearch in a cURL-like syntax. Some features are not included in the basic licence given by Elastic, but require a premium licence purchase: this allows users to explore anomalies in time series data with unsupervised Machine Learning features, which are described in more detail in the following sections.

\section{Managing storage resources on Grid}
The WLGC main mission, since the very beginning, is the capability to share data and resources over a wide-area network across several organizational domains. The current Grid infrastructure, therefore, has to face multiple heterogeneous storage and file system in order to manage, replicate and access files in a distributed system. Moreover, high performance disk-storage solutions have become increasingly important to deal with the large I/O throughput required by the High Energy Physics (HEP) community for both data analysis and Monte Carlo simulations.\\
Given such requirements, the HEP Grid community has developed and implemented the \textit{Storage Resource Manager} (SRM) interface \cite{Storm:SRM}. From one side a client can request the SRM to reserve space and manage files and directories; from the other side, SRM is able to dynamically decide which files to keep inside the storage space and which to remove e.g. to free some disk space. SRM relies on some basic concepts on space and files. A file can be volatile, permanent or durable. A \textit{volatile} file is temporary with a lifetime associated, deleted by a garbage collector of the SRM. A \textit{permanent} file can only be removed by the owner. A \textit{durable} file has a lifetime associated with it, but can be removed by both the garbage collector and the owner.\\
In Grid, a file is identified in several ways: the \textit{Logical File Name} (LFN) is the user friendly name, often user-defined. The \textit{Storage} URL (SURL) points to the physical stored version of a file on a grid storage element. The \textit{Transport} URL (TURL) is an URL issued by SRM for a SURL which can be used to retrieve or store data, e.g. with GridFTP. Finally the \textit{file catalog} is the grid middleware server which maps LFNs to SURLs i.e. tracks the real physical locations of logical file names.\\
Some implementation of SRM service are dedicated to specific storage system e.g. SRM on CASTOR at CERN \cite{SRM:Castor} and SRM on dCache \cite{SRM:dcache}.

\subsection{The StoRM Storage Resource Manager}
\begin{figure}
\centering
\includegraphics[width=0.7\textwidth]{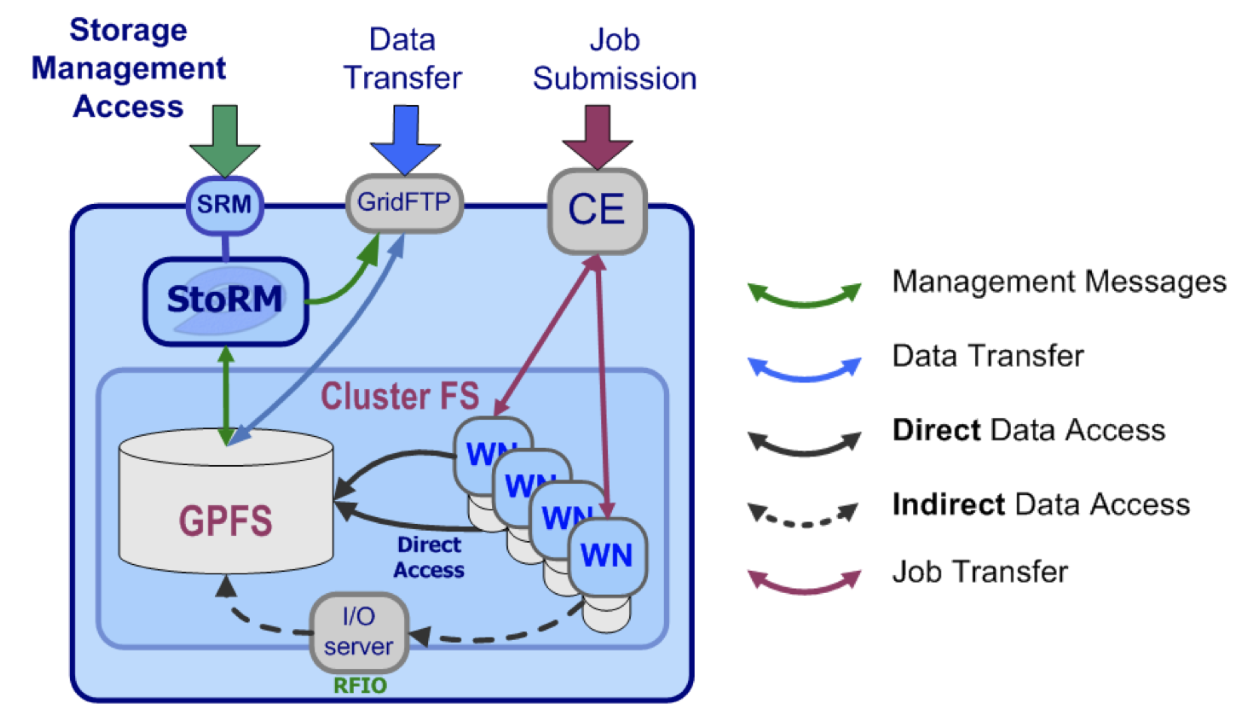}
\caption{StoRM role in a cluster site.}
\label{fig:storm_arch}
\end{figure}

\begin{figure}
\centering
\includegraphics[width=0.6\textwidth]{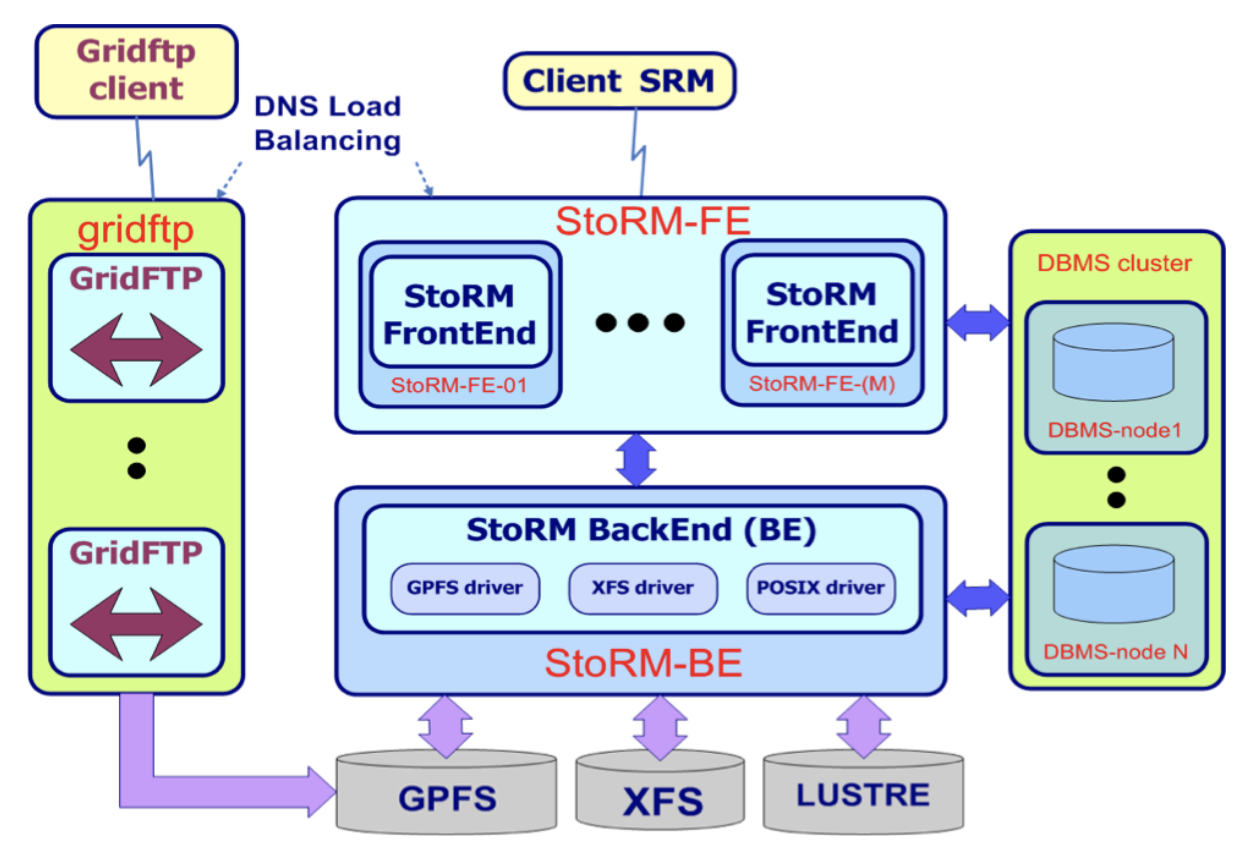}
\caption{Schematic representation of the StoRM architecture.}
\label{fig:storm_arch2}
\end{figure}

StoRM is a Storage Resource Manager that relies on a parallel file system or a standard Posix file system backend i.e. from high performance parallel file systems like GPFS (from IBM) \cite{GPFS:documentation}. StoRM provides advanced SRM functionalities to dynamically manage space and files according to the user requirements, specifying the desired lifetime and allowing for advance space reservation and a different quality of service provided by the underlying storage system. StoRM takes advantage from the file system security mechanisms to ensure access to data in a secure way. Figure \ref{fig:storm_arch} shows the role of StoRM inside a site.\\ StoRM has a multilayer architecture made by two main components: the \textit{front-end} (written in C/C++), and the \textit{back-end} (written in Java).\\
The front-end(s) exposes the SRM web service interface, manages user authentication and stores the requests data into a database. The back-end, instead, is the core of StoRM service, taking care of executing all the synchronous and asynchronous SRM functionalities. It is responsible of the management of file and space metadata, it enforces authorization permissions on files and it interacts with other Grid services. Moreover, the back-end is able to use advanced functionalities provided by the file system (for example by GPFS and XFS) to accomplish space reservation requests.\\
Currently, StoRM is adopted in the context of WLCG infrastructure in various data centers, including the Italian Tier 1 at the CNAF-INFN institute. In order to satisfy the high availability and scalability requirements coming from the HEP community, StoRM can be deployed in a clustered configuration, with multiple instances of front-end and back-end services and with a dedicated DBMS installation: multiple front-end instances can be deployed on separate machines and configured to work on the same database and the same back-end service.
In Figure \ref{fig:storm_arch2}, a sketch of this architecture is shown.

\section{StoRM Logging information}
StoRM services produce a huge quantity of log information, with different levels of verbosity that are stored in different files.

\subsection{StoRM Frontend Logging}
The Frontend (FE) logs information on the service status and SRM requests received and managed by the process. Different levels of logging are supported such as ERROR, WARNING, INFO, DEBUG that can be set from dedicated parameters in a configuration file. 

\subsubsection*{storm-frontend-server.log}
\begin{figure}
\centering
\includegraphics[width=\textwidth]{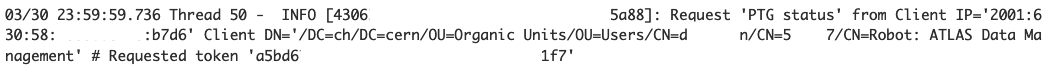}
\caption{\textit{storm-frontend-server} log line example.}
\label{fig:frontend_server_log}
\end{figure}

The Frontend log file named \textit{storm-frontend-server.log} is placed in the /var/log/storm directory. At start-up time, the FE prints here the whole set of configuration parameters, this can be useful to check desired values. When a new SRM request is managed, the FE logs information about the user subject name, community group and role identifiers as well as several other details of the request parameters, as shown in Figure \ref{fig:frontend_server_log}.

\subsubsection*{monitoring.log}
\begin{figure}
\centering
\includegraphics[width=\textwidth]{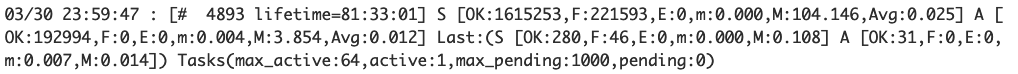}
\caption{\textit{monitoring} log line example.}
\label{fig:monitoring_log}
\end{figure}

The monitoring service, if enabled, provides information about the operations executed within a certain time frame, writing them to file as shown in Figure \ref{fig:monitoring_log}. This amount of time (called Monitoring Round) is configurable and its default value is 1 minute. At each Monitoring Round, a single row is printed in the log. This row reports both information about requests that have been performed in the last Monitoring Round and information considering the whole FE execution time (Aggregate Monitoring). Information reported are generated from both Synchronous (operations that return the control to the client when the request has been executed) and Asynchronous requests (that return the control as soon as the request has been accepted by the system) and tell: how many requests have been performed in the last Monitoring Round, how many of them were successful, how many failed, how many produced an error, the average execution time, the minimum execution time and the maximum execution time.

\subsection{StoRM Backend Logging}
The Backend (BE) log files provide information on the execution process of all SRM requests. Backend logging is based on \textit{logback} framework. Logback provides a way to set the level of verbosity depending on the use case. The level supported are FATAL, ERROR, INFO, WARN, DEBUG.

\subsubsection*{storm-backend.log}
\begin{figure}
\centering
\includegraphics[width=\textwidth]{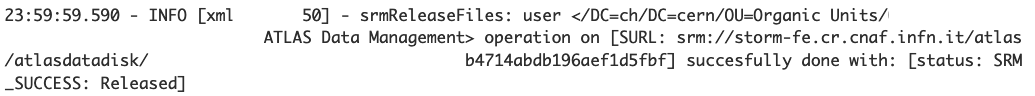}
\caption{\textit{storm-backend} log line example.}
\label{fig:backend_log}
\end{figure}

This log file contains all the information about the SRM execution process, error or warning. At start-up time, the BE logs here all the properties value, which can be useful to check parameters effectively adopted by the system. After that, at the INFO level, the BE logs for each SRM operation who has requested the operation (user subject name), on which files (SURLs) and with which result. If ERROR or FATAL levels are set, the only event logged in the file are those due to error conditions. In Figure \ref{fig:backend_log} an example of \textit{storm-backend} log is shown.

\subsubsection*{heartbeat.log}
\begin{figure}
\centering
\includegraphics[width=\textwidth]{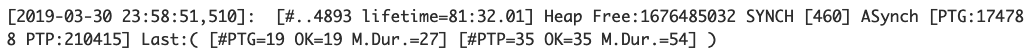}
\caption{\textit{heartbeat} log line example.}
\label{fig:heartbeat_log}
\end{figure}

StoRM provides a bookkeeping framework that elaborates informations on SRM requests processed by the system to provide user-friendly aggregated data that can be used to get a quick view on system health. The \textit{heartbeat.log} file contains information on the SRM requests processed by the system from its startup, adding new information at each beat. The beat time interval can be configured, by default is 60 seconds. At each beat, the heartbeat component logs an entry.\\
As shown in Figure \ref{fig:heartbeat_log}, the information contained in this line are the lifetime from the last startup, the BE process free heap size in Bytes, the number of Synchronous SRM requests executed in the last beat, the number of \textit{srmPrepareToGet} and \textit{srmPrepareToPut} requests executed from start-up and the number of \textit{srmPrepareToGet} (as well as the \textit{srmPrepareToPut}) executed in the last beat, with the number of requests terminated with success (OK=10) and the average completion time in millisecond (M.Dur.=150).\\
This file can help you to understand if the system is receiving SRM requests, or if the system is overloaded by SRM requests, or if PTG and PTP are running without any problem, or if the interaction with the filesystem is exceptionally slow (in case the mean duration is much higher than usual).

\subsubsection*{storm-backend-metrics.log}
\begin{figure}
\centering
\includegraphics[width=\textwidth]{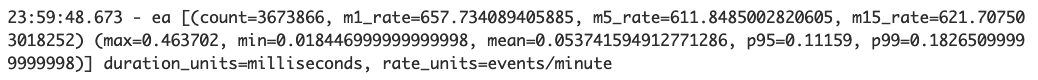}
\caption{\textit{storm-backend-metrics} log line example.}
\label{fig:backendmetrics_log}
\end{figure}

A finer grained monitoring of incoming synchronous requests is provided by this log file, which contains metrics for individual types of synchronous requests. An example entry log of \textit{storm-backend-metrics.log} is shown in Figure \ref{fig:backendmetrics_log}: the information stored are the type of operation, the number of operation in the last minute, the number of operations from last startup, the maximum (minimum and average) duration of last bunch and the highest duration of the 95\% (and 99\%) last bunch operations.

\section{Building the infrastructure with the Elastic Stack}
The installation and setup of the Elastic Stack suite has been performed on a test-bed environment in the INFN-CNAF data center using a VM in Openstack \cite{OPENSTACK:doc}. In particular the specifics of the VM test-bed used are the following:
\begin{itemize}
\item 2.2 GHz Dual Core VCPUs (Broadwell architecture) with 4MB cache;
\item 4 GB of memory;
\item 40 GB of disk;
\item two volumes attached with a cumulative storage of 600 GB
\end{itemize}

\subsection{Parsing the log information}
\begin{figure}
\centering
\includegraphics[width=\textwidth]{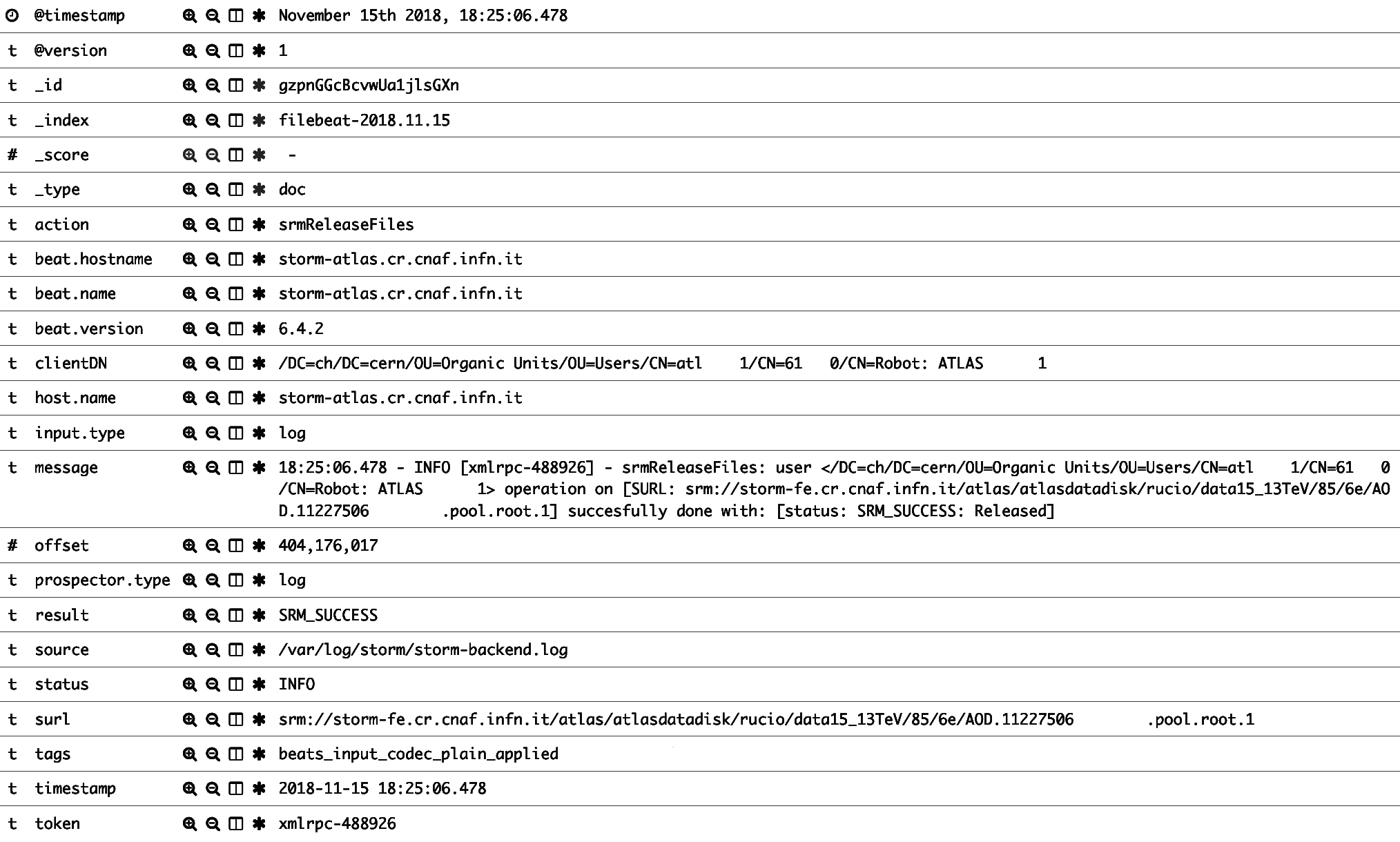}
\caption{Example of StoRM Backend metrics log line parsing. }
\label{fig:parsed}
\end{figure}

The information coming from the log sources, described in the previous section, contains heterogeneous data that require a structured filtering in order to be correctly indexed. Inside the local cluster itself, Logstash allows the creation of a well defined pipeline collecting data from Filebeat, which is installed in each node that contains the log files.\\
The different choice of filters is essential in order to correctly parse the log content: in particular, the \textit{grok} filter has been adopted. A \textit{grok} filter, based on Regular Expressions, selects a specific portion of text (both numeric or literal depending on the information required) by creating a series of patterns. Some patterns are predefined, such as the IP address or the timestamp ISO format (which are universally adopted and recognized), but the majority are custom made and stored in a specific configuration file.\\
In Figure \ref{fig:parsed} an example of parsed log, with new structured information, is shown. In this example, the original message is contained inside the \textit{message} field and the other information parsed from the initial log are listed alphabetically, such as the \textit{status} of the StoRM operation (INFO) or the \textit{action} containing the specific operation performed (\textit{srmReleaseFiles}). Some fields are automatically added by Filebeats, e.g. the information of the sender node name (\textit{beat.name}), the index name to which the log belongs, the \textit{offset} field corresponding to the file offset the reported line starts at and the type of document processed (\textit{log} in this case).

\subsection{Data visualization}
\begin{figure}
\centering
\includegraphics[width=0.84\textwidth]{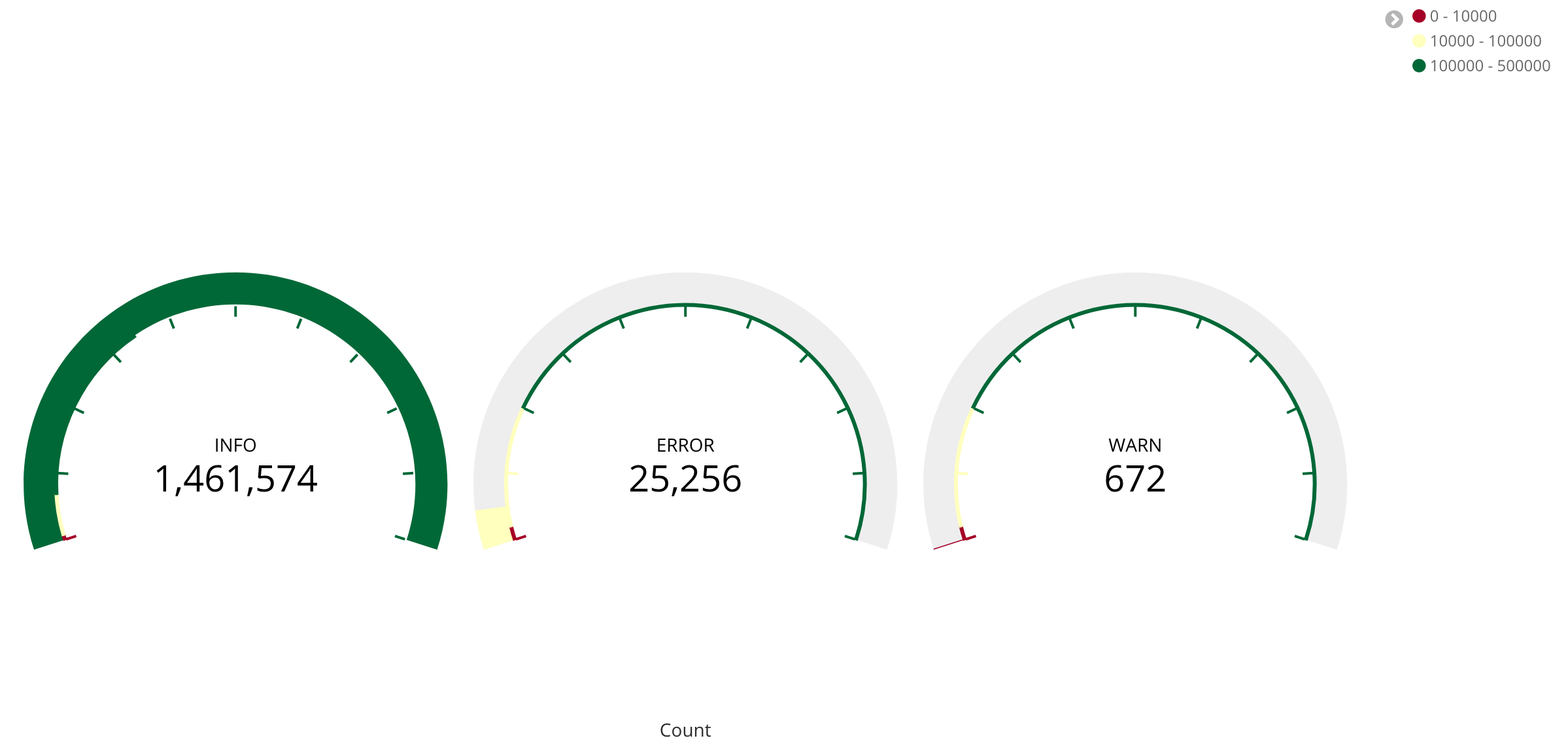}
\caption{Gauge visualisation with the possible statuses of the StoRM Backend.}
\label{fig:gauge}
\end{figure}

\begin{figure}
\centering
\includegraphics[width=\textwidth]{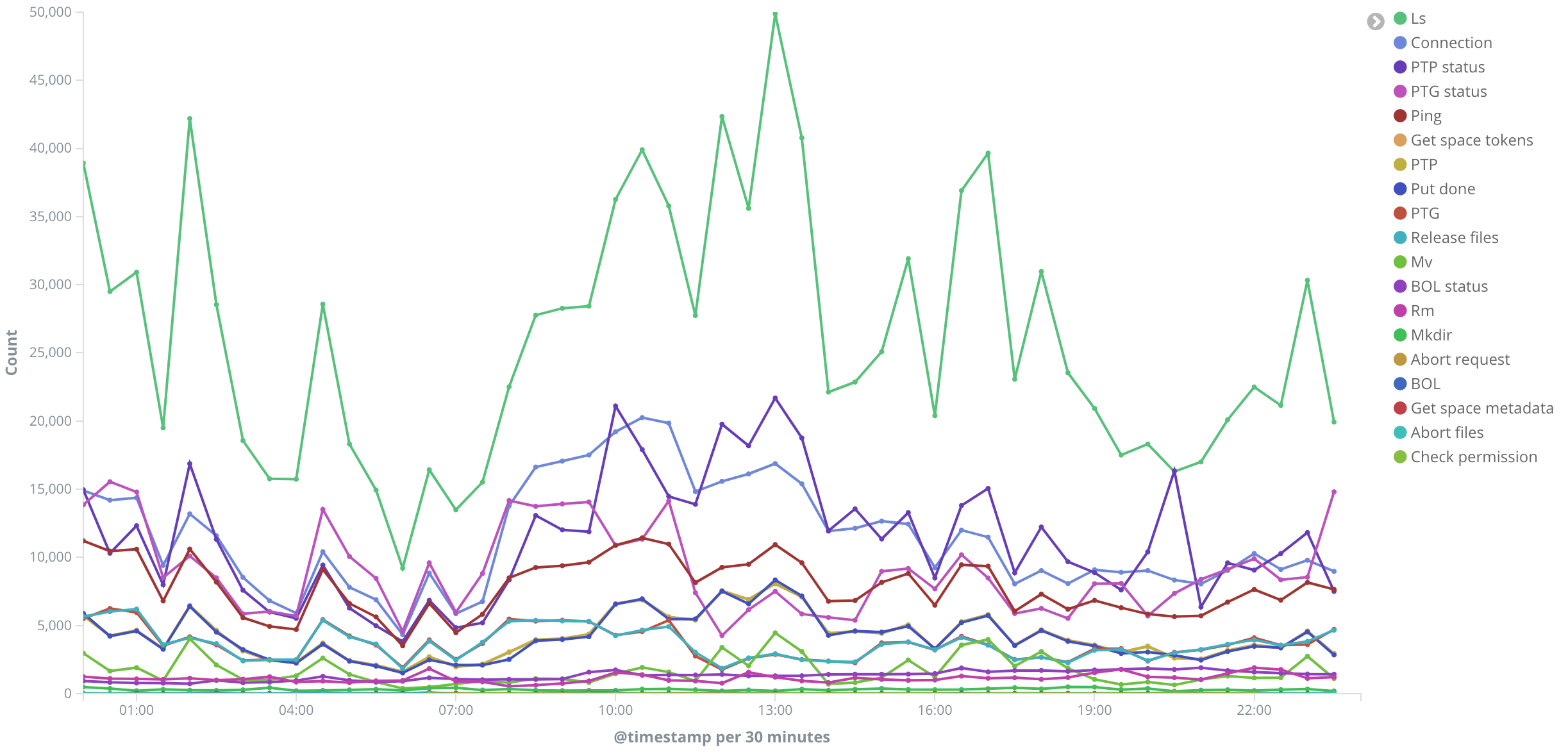}
\caption{Count of different requests for the StoRM Frontend.}
\label{fig:requests}
\end{figure}

\begin{figure}
\centering
\includegraphics[width=0.8\textwidth]{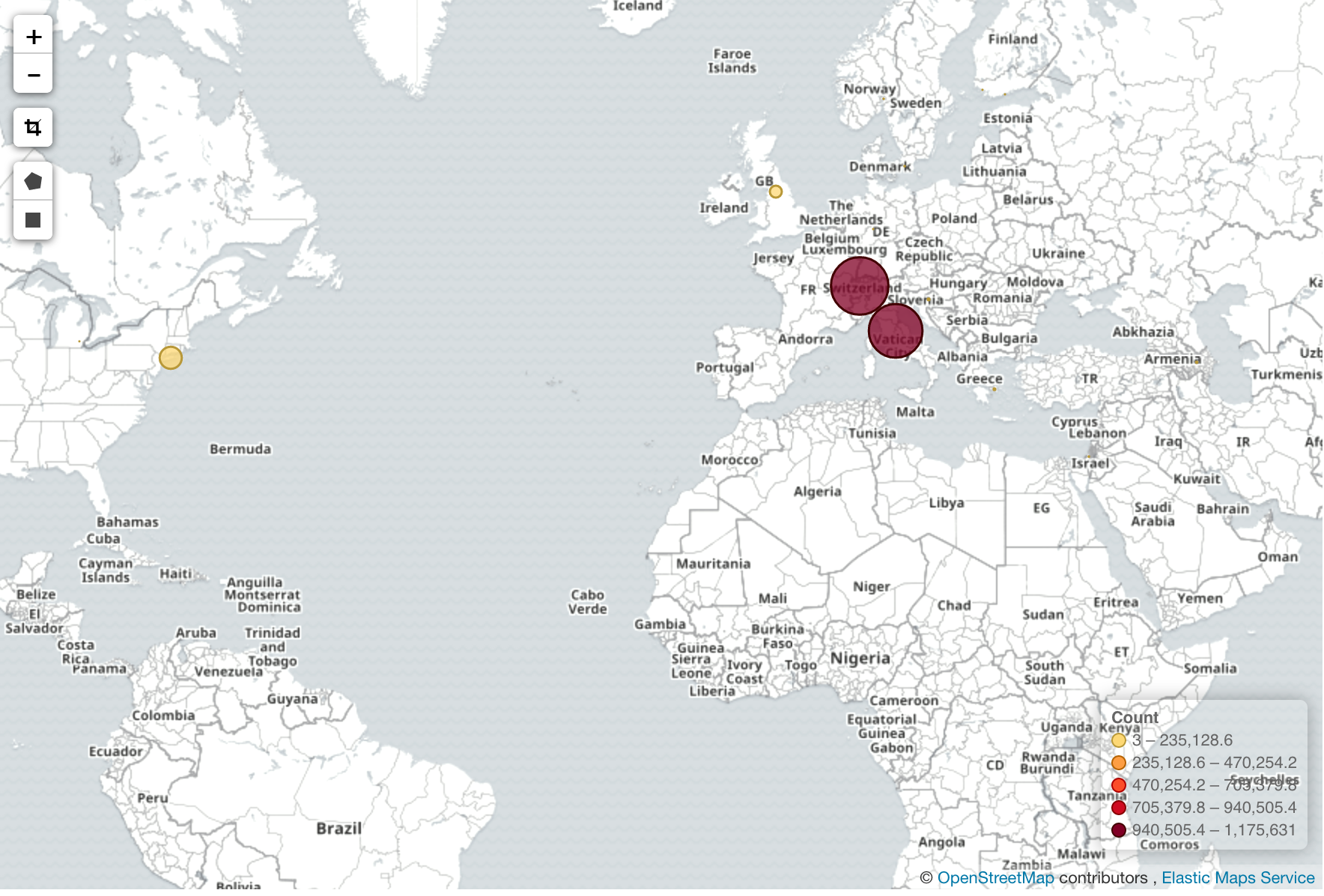}
\caption{Geo heat map for the IP address location of the most frequent StoRM requests.}
\label{fig:map}
\end{figure}

Using the Kibana User Interface, the information extracted from the log parsing is then processed and plotted with several histograms and charts. In Figure \ref{fig:gauge}, a gauge chart shows the count of the different status in the StoRM Backend log file on a specific time range defined by the user. Another possible plot, shown in Figure \ref{fig:requests}, represents the count of all possible StoRM Frontend operations in a time series format showing the user's most frequent (Ls, Connection, status of Prepare To Get and Prepare To Put). Using a heat map view, it is also possible to geographically locate the most frequent IP addresses of StoRM client requests, as shown in Figure \ref{fig:map}.

\section{Machine Learning with the Elastic Stack}
\begin{figure}
\centering
\includegraphics[width=\textwidth]{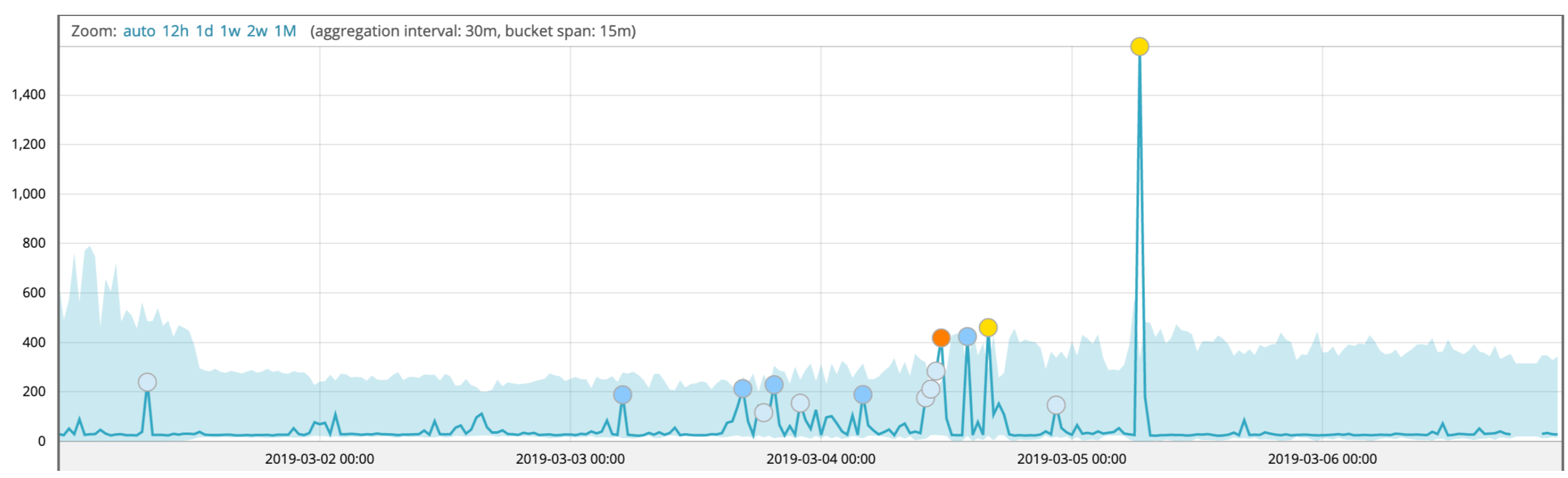}
\caption{Time series anomaly detection for the last bunch of \textit{ptg} operations, from \textit{heartbeat} log.}
\label{fig:predictive0}
\end{figure}

\begin{figure}
\centering
\includegraphics[width=\textwidth,height=0.18\textheight]{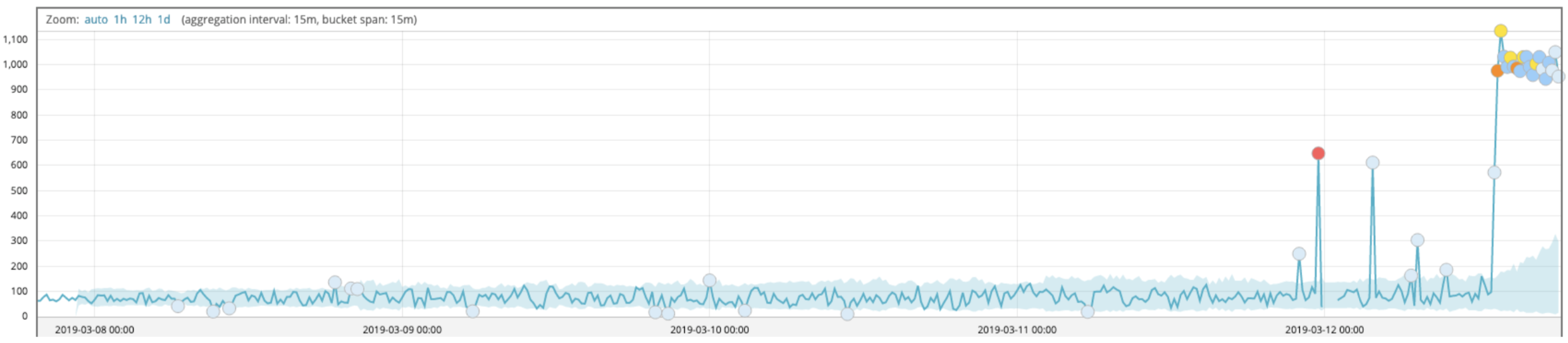}
\caption{Time series anomaly detection for the mean duration of the continuous last bunch of operations, from \textit{storm-backend-metrics} log.}
\label{fig:predictive1}
\end{figure}

In the Elasticsearch new release (version 6.0 and above), a new extension was released using Machine Learning capabilities for data search and analytics. This functionalities are embedded within the premium features of \textit{X-Pack} and are mainly focused on the time series anomaly detection. Using unsupervised proprietary algorithms, the most straightforward use case of this technology is to identify when a metric value or event rate deviates from its "normal" behaviour. Therefore, the entry point into the Machine Learning features is a Single Metric job that allows the investigation of data trend during normal activity in order to identify anomalies on univariate time series data. If the anomalies found are useful, it is then possible to run the analysis continually in real time and alert when an anomaly occurs.\\
The implementation of such functionalities is optimized to run in an Elasticsearch cluster, allowing the elaboration of millions of events in a small amount of time. An example is shown in Figure \ref{fig:predictive0}, using the information from the last continuous bunch of \textit{srmPrepareToGet} synchronous operations from the \textit{heartbeat} log file. The light blue area corresponds to the model bounds created, at first, by the training phase of a past time interval and then by the real-time training process. Data coming from the log is shown as a straight line and - during a normal activity - should be contained inside the model bounds; when an anomaly occurs, the metric value exceeds the model threshold and, therefore, produces a warning alert with a severity proportional to their relative discrepancy.\\
Figure \ref{fig:predictive1}, instead, shows an application of these functionalities with data coming from the parsed information of the StoRM Backend metrics log at the CNAF data center. Using the mean duration of the last bunch of operations described in each log entry, it is possible to notice that - starting at a certain point - the time duration increases by an order of magnitude causing, therefore, an increase also in the number of warnings (shown in figure as small coloured dots based on the level of alert). Figure \ref{fig:predictive1} also depicts another important consideration: in fact - in the few hours preceding the sudden increase in the metric value - some anomalous isolated peaks may indicate and may be associated with the subsequent failure. The proactive identification of the anomaly, however, is not possible using this particular tool given in the Elastic Stack. The only tool provided by the Machine Learning functionality consists in a \textit{forecast} button which purpose is the prediction of a time interval, given a specific portion of training previous data. This option, however, does not take under consideration any possible quick fluctuation of the examined metric since only the average behaviour of the trained model is used for making future predictions. 

\subsection{Computing resources management} 
\begin{figure}
\centering
\includegraphics[width=\textwidth, height=0.3\textheight]{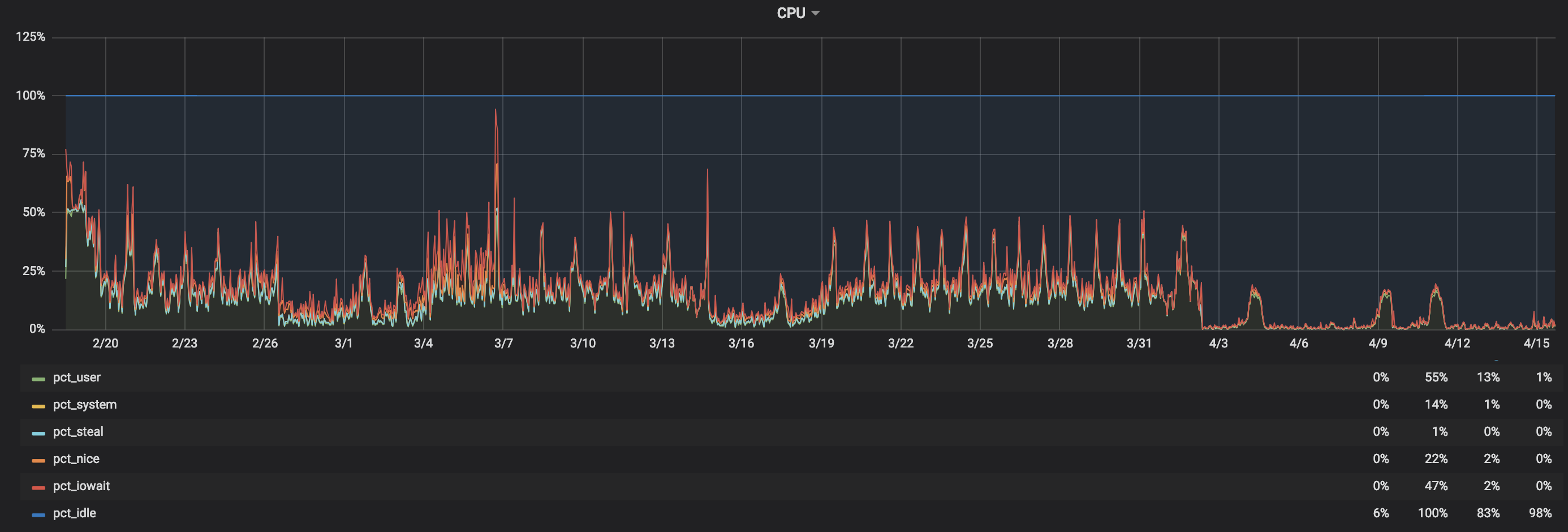}
\caption{CPU usage on a two months time range.}
\label{fig:cpu}
\end{figure}

\begin{figure}
\centering
\includegraphics[width=\textwidth, height=0.3\textheight]{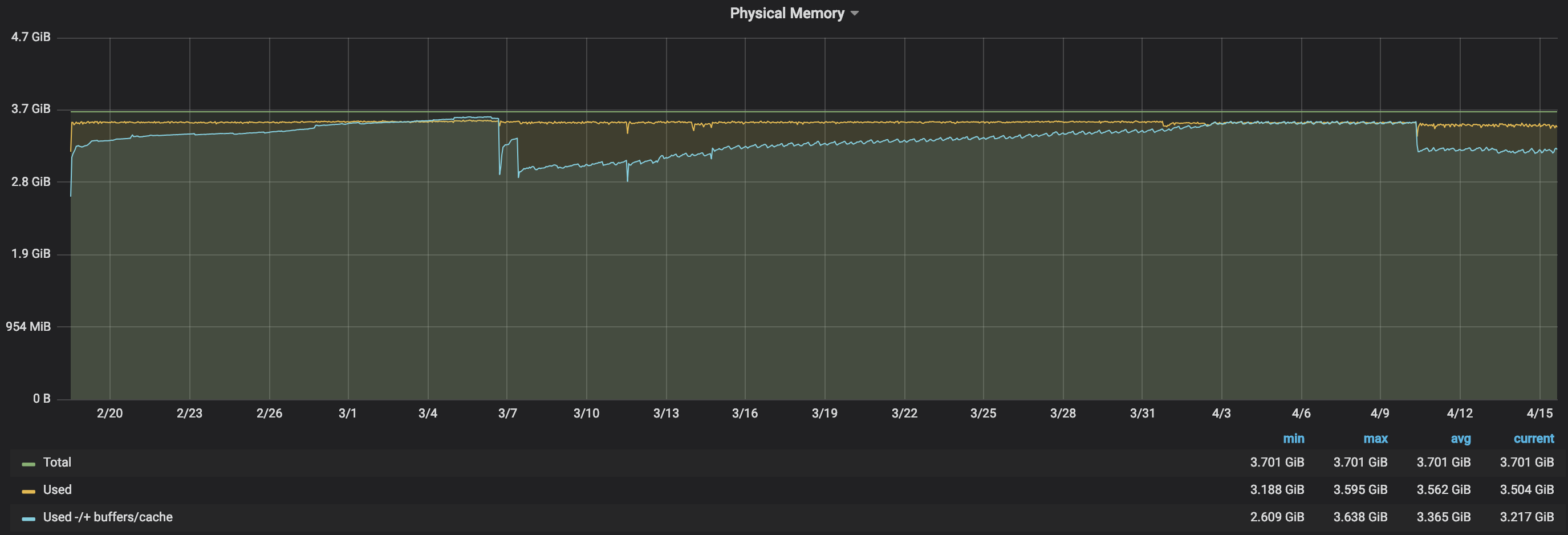}
\caption{Memory usage on a two months time range.}
\label{fig:memory}
\end{figure}

\begin{figure}
\centering
\includegraphics[width=\textwidth, height=0.3\textheight]{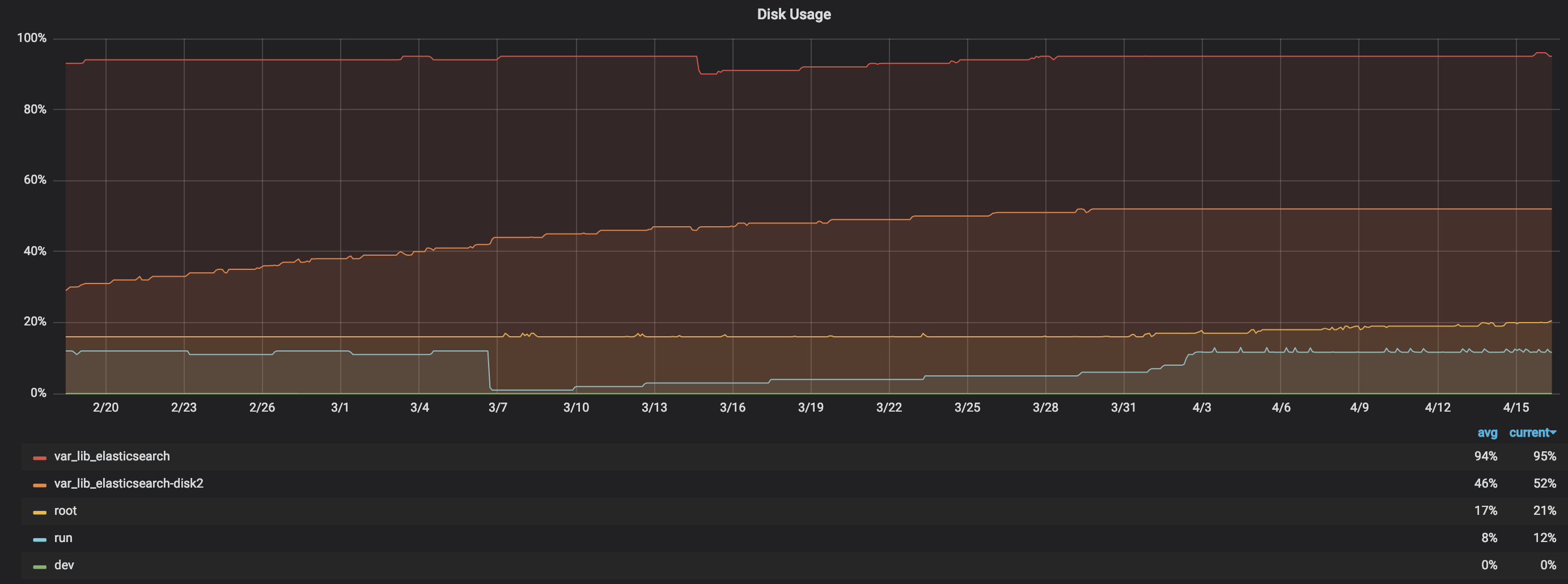}
\caption{Storage usage on a two months time range.}
\label{fig:storage}
\end{figure}

Another important aspect to consider is the workload on the VM resources used by the Elastic Stack test-bed environment itself. The performance of the computing resources provided are, in fact, continuously monitored and displayed using Graphana \cite{Graphana:doc}.\\
In Figure \ref{fig:cpu}, the behaviour of the CPU usage over time is shown: a fair amount of CPU utilization is observed in the time range shown, even during the most intensive operations reaching a maximum value of about 90\% but with an average value of 35-40\%. The predominant task involving CPU time is the real-time Machine Learning anomaly detection analysis (over several different jobs submitted, using different test metrics); the main indication of such hypothesis is the sudden drop in CPU in the last quarter of the above-mentioned chart caused by the termination of each job due to licence expiration.\\
Memory usage, instead, is critical using the Elastic Stack as shown in Figure \ref{fig:memory}: during the entire time range, in fact, memory is always fully saturated. The only noticeable decrease - visible in figure - is mainly caused by the manual interruption of older indexes in order to optimize memory usage.\\
Concerning storage, two volumes are used: the first one, which is currently full, of 200GB and a second one, currently in use, of 400GB. Figure \ref{fig:storage} shows the filling trend of the second disk (orange line) in the two months' observation window.

\section{Conclusion}
The Elastic Stack suite is a powerful tool for log analysis and ingestion as well as their indexing and visualisation for online monitoring. At the INFN-CNAF Tier-1 data center, this suite has been investigated for the creation of a centralised platform for logs coming from the StoRM service.\\
Using a test-bed environment on an isolated virtual machine, log files coming from different StoRM machines were collected and parsed using specific filters based on Regular Expressions. In this way, the information carried out by such logs is structured and indexable by the Elasticsearch engine providing fast search capabilities and visualisation of specific variables.\\
Moving towards a predictive failure system, the Elastic Stack suite provides a premium functionality with a Machine Learning approach which was investigated and tested. This system, using unsupervised learning techniques, is mainly used for an anomaly detection scenario by alerting operators when a certain metric exceeds a specified threshold given by the trained model, continuously updated with new information. Using several metrics coming from the StoRM Backend and Frontend instances during anomalous observation windows, several alerts were correctly identified by the system. However, for a predictive scenario and a proactive identification of failures, this may not be the optimal solution.\\
As a short-term goal at the INFN-CNAF computing center, some plans are currently being investigated by developers. The first step is the creation of a centralised log source stored inside a physical storage partition of the Tier-1. This unique log file will contain all the information appended from different services running and may be accessed from any virtual machine by a NFS mount point.
A clean installation for the Elastic Stack suite will then be created on a dedicated physical cluster and - using the Elasticsearch engine to process all the log information - adopt other Machine Learning algorithms using conventional frameworks \cite{giommi:paper}.\\
Another possible direction towards Big Data processing is also taken under consideration by adding an Apache Spark \cite{spark:apache} cluster on a cloud machine at CNAF with three dedicated storage volumes of about 300GB each (installed via DODAS@CNAF).\\
Finally, also other logs must be analysed as well: Worker Nodes status, service machines, GPFS, GridFTP \cite{gridftp:doc}, xrootd \cite{xrootd:doc}, batch system and application level logging.

\end{document}